\documentclass[aps,twocolumn,nofootinbib,superscriptaddress,showpacs]{revtex4}
\usepackage{graphicx}
\usepackage{amsfonts}
\usepackage{psfrag}
\usepackage{amsmath}
\usepackage{float}
\usepackage{amssymb}  

\newcommand{\van}{\scriptstyle}

\def\be{\begin{equation}}
\def\ee{\end{equation}}
\def\ba{\begin{eqnarray}}
\def\ea{\end{eqnarray}}

\def\Tr{\text{Tr}}

\def\ut#1{\rlap{\lower1ex\hbox{$\sim$}}#1{}}

\DeclareFontFamily{U}{rsfs}{}         
\DeclareFontShape{U}{rsfs}{m}{n}{<5> rsfs5 <6><7> rsfs7          %
  <8><9><10><10.95><12><14.4><17.28><20.74><24.88> rsfs10}{}     %
\DeclareMathAlphabet{\mathfs}{U}{rsfs}{m}{n}                     %
\newcommand{\mfs}[1]{\mathfs {#1}}                               %

\newcommand{\sM}{{\mfs M}}

\newcommand{\sI}{{\mfs I}}

\begin{document}

\title{No firewalls in quantum gravity: the role of discreteness of quantum geometry in resolving the information loss paradox}

\author{Alejandro Perez}
\affiliation{
    Aix Marseille Université, CNRS, CPT, UMR 7332, 13288 Marseille, and
    Université de Toulon, CNRS, CPT, UMR 7332, 83957 La Garde, France.
}

\begin{abstract}
In an approach to quantum gravity where space-time arises from coarse graining of fundamentally discrete structures, black hole formation and subsequent evaporation can be described by a unitary evolution without the problems encountered by the standard remnant scenario or the schemes where information is assumed to come out with the radiation while evaporation (firewalls and complementarity).  The final state is purified by correlations with the fundamental pre-geometric structures (in the sense of Wheeler) which are available in such approaches, and, like defects in the underlying space-time weave, can carry zero energy.    \end{abstract}


\maketitle
\section{Introduction}

The second law of thermodynamics is not a fundamental principle in physics but rather a statement
about the (illusory) asymmetry of time evolution when sufficiently complicated systems are put in special initial conditions and later described statistically in terms of coarse physical variables that are unable to discern all the details of the fundamental system. The idea is easily illustrated in classical mechanics. On the one hand, Liouville's theorem implies that the support of the phase space distribution of the system spans a volume that is time independent; on the other hand, the shape of the support is not restricted by the theorem. An initially simple distribution supported in a ball in $\Gamma$ will (in suitably complicated systems) evolve into a more an more intricate shape whose apparent phase space volume, when measured with a devise of resolution lower than that of the details of the actual distribution, will grow with time.    

In a practical sense, the second law implies that information is degraded (yet not lost) in time when encoded in 
coarse variables. The words in a newspaper are gone when the newspaper is burned, the data in a hard drive deteriorate due to fluctuations in the magnetic medium, the writings on a marble headstone wash out under the obstinate action of rain drops. In some cases information can be long lived if encoded in 
variables protected from the statistical fluctuations of the ambient medium by an energy threshold 
(e.g. binding energy of DNA molecules at room temperature) or, in a more quantum mechanical setting, by the existence of decoherence free subspaces (e.g. in quantum computing systems). However, as in the opening examples of this paragraph, this information will be degraded as soon as the ambient conditions are suitably changed so that 
new degrees of freedom become available to the system. In all these examples there is a physical account of the phenomena where information is preserved in the sense that the system is unitary. Information goes into the correlations among more basic degrees of freedom which are inaccessible to the (coarse) observer that defined the notion of information in the first place: after burning the newspaper the story remains written in the correlations of the gas molecules diffusing in the atmosphere. 

In this article we argue that, as in the previous more familiar situations, information is degraded but not lost in black holes. The underlying quantum discreteness
of geometry, expected from non perturbative quantum gravity approaches, provides a simple mechanism for maintaining unitary evolution without running into the contradictions recently raised in \cite{Braunstein:2009my, Almheiri:2012rt}, and without radically changing the semiclassical understanding of black hole dynamics \cite{Rovelli:2014cta, Haggard:2014rza}. However, as in the more familiar systems mentioned above, and despite unitarity of evolution,  the underlying quantum geometry degrees of freedom imply the degradation of low energy quantum field theoretical  information when black holes form and subsequently evaporate due to the standard phenomenon of decoherence. 

As a consequence, we will argue that, the usual scattering or $S$-matrix approach based on an background geometry in asymptotic regions cannot be viable for the description of the fundamental unitary evolution.  According to the view advocated here,  such formulations are bound to miss relevant correlations with degrees of freedom that are not describable as field excitations on a smooth background geometry and will thus run into problems with unitarity.  The fundamental system is unitary but cannot be described in terms of fields living on a background geometry. 
One recovers in this sense the original Hawking conclusion \cite{Hawking:1976ra} as far as low-energy QFT degrees of freedom living on a classical geometry is concerned.

\subsection{The semiclassical global picture}

In the semiclassical regime, black holes are thermodynamical systems close to thermal equilibrium.
The singularity theorems, together with the cosmic censor conjecture, and the no hair theorem, imply that (classically) stationary black holes are the final result of an isolated gravitational collapse. This is the first indication of the irreversible character of gravitational collapse: an infinite dimensional set of initial conditions 
is expected to lead to a final state with a  black hole described by only three macroscopic parameters mass $M$, angular momentum $J$ and electric charge $Q$.  

However, this final state is not stable if quantum effects are not neglected. According to the semiclassical analysis (or gravitational mean field approximation) black holes slowly evaporate due to vacuum fluctuations by emitting radiation with a thermal spectrum of temperature $T_{H}=\kappa/(2\pi)$ where $\kappa$ is the surface gravity. This radiation carries energy (as well as charge and angular momentum) away to future null infinity $\sI^+$ and its back reaction causes the mass of the black hole (and its angular momentum and charge) to decrease. The semiclassical picture breaks down when the BH becomes small as the curvature at the horizon approaches Planck's curvature. The final dynamical stages of evaporation need a full quantum gravity treatment.  

\begin{figure}[h] \centerline{\hspace{0.5cm} \(
\begin{array}{c}
\psfrag{Ip}{$\sI^{+}$}\psfrag{Im}{$\sI^{-}$}
\psfrag{i0}{$i_0$}\psfrag{ip}{$i^{+}$}\psfrag{im}{$i^{-}$}
\psfrag{s}{$$}
\psfrag{S1}{$\Sigma_1$}
\psfrag{S2}{$\Sigma_2$}
\includegraphics[width=5cm]{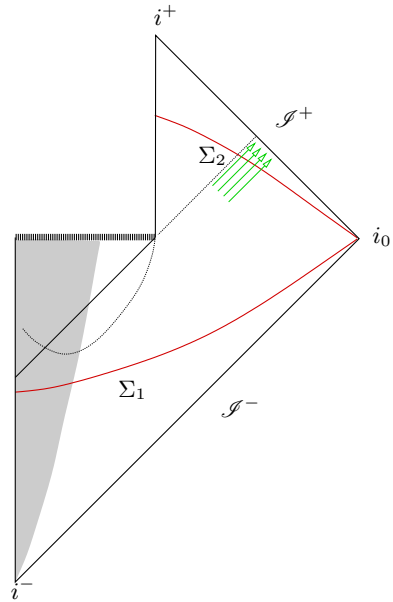}
\end{array}
\)}
\caption{The standard semiclassical global causal structure representing black hole formation and the subsequent evaporation.  The hyper surface $\Sigma_2$ representing an `instant' of time after the complete evaporation of the initial black hole, fails to be a Cauchy surface of the whole space-time.} \label{standard}
\end{figure}

In his seminal paper Hawking \cite{Hawking:1976ra}  considered the situation illustrated in Fig. \ref{standard}, where the Carter-Penrose diagram depicting the whole history of black hole formation and evaporation is given. The curved dotted line denotes the initially growing apparent horizon (due to the infall of collapsing matter and/or gravitational radiation) that subsequently shrinks until disappearing (due to the back reaction of the Hawking radiation).  Such dynamical features of apparent horizons are well described by the notion of {\em dynamical horizons} (see \cite{AKLR} and references therein). The BH horizon does not coincide with the apparent horizon, but shares with it the initial rapid (classical) phase of growth followed by the (quantum) phase of slow shrinking due to the Hawking effect. Hawking radiation, represented by the arrows crossing the hypersurface $\Sigma_2$, is approximately thermal for the very long time period while the black hole horizon is large in Planck units. More precisely, the retarded time span $\Delta u$ of the evaporation process, as measured by observers in $\sI^{+}$, is $\Delta u \approx M^3/\ell_p^2$ where $M$ is the initial Bondi mass of the BH. During this time, the Bondi mass $M_B(u)$ of the systems decreases from the initial value M to essentially zero. This means that the space-time is (essentially) flat once an observer at $\sI^+$ crosses the Cauchy horizon represented by the dotted line prolonging the BH horizon to $\sI^+$ in the figure.

If Fig. \ref{standard} represents the correct global causal structure of the system
it is then clear that one cannot have a conventional notion of unitary evolution 
from an initial instant of time $\Sigma_1$ to a later instant of time $\Sigma_2$.
The reason is that in contrast to $\Sigma_1$,  the hypersurface $\Sigma_2$ is not a Cauchy surface.
This means that an hypothetical pure state of the system defined on $\Sigma_1$ 
will have to be represented by a mixed state on $\Sigma_2$ due to the fact that 
some degrees of freedom on $\Sigma_1$ cannot be reconstructed from measurements on $\Sigma_2$ as they  are lost when falling into the singularity.
As emphasized by Wald \cite{lrr-2001-6} there is no fundamental problem with this loss of information
which can indeed take place for very similar reasons on flat space-time examples \cite{Wald}.
Notice that the loss of unitarity is just a direct consequence of the causal structure of an evaporating BH whose key features only depend on the fact  that  quantum effects allow for a positive energy flow at $\sI^+$. In particular, the degree to which Hawking radiation is thermal or not has no importance for the above discussion\footnote{By Hawking radiation here we mean the black hole radiation detected at $\sI^+$ from the retarded time where the BH has become stationary with an initial Bondi mass $M\gg M_p$ to the retarded time where its mass has reduced to the order of $M_p$. This is the era of the system that we assumed to be well described by the semiclassical treatment.}.  

The structure of space-time in Fig. \ref{standard} led Hawking to conclude that time evolution, in the context of gravitational collapse, could not be captured by the usual notion of scattering $S$ matrix. As pure states on $\Sigma_1$ will necessarily evolve into density matrices on $\Sigma_2$, Hawking postulated a new framework where time evolution is given by a suitably defined super-scattering operator  evolving density matrices into density matrices \cite{Hawking:1976ra}.  Hawking proposal was criticised by Banks, Peskin, and Susskind  \cite{Banks:1983by} on the basis that it could lead to strong conflicts with energy-momentum conservation. However,  Wald and Unruh \cite{Unruh:1995gn} exhibit examples satisfying Hawking's postulates that easily avoid such inconsistency claims by invoking decoherence with hidden energy cheap degrees of freedom (our scenario be similar in spirit to these examples).

\begin{figure}[h] \centerline{\hspace{0.5cm} \(
\begin{array}{c}
\psfrag{Ip}{$\sI^{+}$}\psfrag{Im}{$\sI^{-}$}
\psfrag{i0}{$i_0$}\psfrag{ip}{$i^{+}$}\psfrag{im}{$i^{-}$}
\psfrag{s}{$$}
\psfrag{S1}{$\Sigma_1$}
\psfrag{S2}{$\Sigma_2$}
\includegraphics[width=5cm]{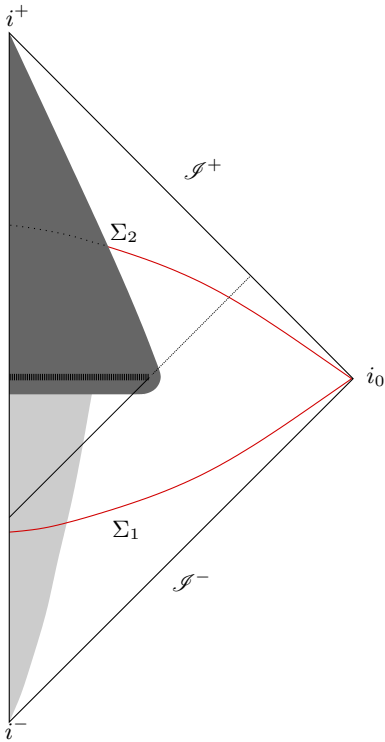}
\end{array}
\)}
\caption{Global causal structure in the remnant paradigm. The darken region represents the region where quantum gravity effects are expected to be large. The black hole evaporates following the semiclassical (mean field) expectation until it becomes Planckian in size. Most of the initial mass is radiated as Hawking radiation and one ends up with a Planckian mass remnant moving along a time-like world-tube of Planck size in (essentially) flat Minkowski space-time with a very long lifetime. Here we assume the remnant is stable so it goes up to $i^+$.} \label{remnant}
\end{figure}

 A concrete proposal of such type consists of assuming that a {\em remnant} of a mass of the order of Planck mass at the end of the Hawking evaporation can carry the missing information \cite{Aharonov:1987tp, Giddings:1992hh}.
As the final phase of evaporation lies outside of the regime of applicability of the semiclassical
analysis such hypothesis is in principle possible. The global space-time causal structure of the remnant proposal is represented in Fig. \ref{remnant}. Strong quantum gravity effects are important in the darken region close around and to the future of the {\em would be singularity}. There fluctuations of geometric quantities might be so large that the usual space-time representation 
cannot be applied. We are assuming that causal relations are still meaningful which is a necessary condition for any discussion of unitary evolution. Late observers at instants of time such as $\Sigma_2$ intersect the quantum gravity region along a time-like world-tube of radius of the order of the Planck length and mass of the order of Planck mass. This is the work-tube of the remnant which looks as a point particle to outside observers.

In order to purify the state at $\Sigma_2$ the remnant must have a huge number of internal degrees of freedom which correlate with those of the radiation emitted during evaporation. If one traces out these degrees of freedom one has a mixed state that represents well the physics of local observers at $\sI^+$. The entropy of such mixed state is expected to be at least as big as the one of the initial black hole $S_{BH}(M)$ before Hawking evaporation starts being important. The value of $S_{BH}(M)$ is a lower bound of the number of such internal states.
This huge internal degeneracy leads to the expectation of an (unobserved) very large pair production rates in standard particle physics situations if one thinks of the remnant as a residual particle.

In Fig. \ref{remnant} we have assumed that the remnant is stable and so its world-tube gets all the way up to $i^+$. One can contemplate the possibility that they could decay via emission of soft photons carrying the missing information back to $\sI^+$. As the energy available for this is of the order of $M_p$,  remnants would have to be very long lived (with lifetimes of the order of $M^4/M_p^4 t_{p}$) in order to evacuate all the internal information in electromagnetic radiation.  Hence, they would basically behave as stable particles and one would run into the previous difficulties with large pair creation rates. The possibility that such remnants  can lead to finite rate production despite of the large degeneracy of their spectrum has been suggested \cite{Banks:1992is, Banks:1992mi}. More discussion of remnants and references see \cite{Hossenfelder:2009xq}.

\subsection{The BH complementarity and the firewall problem}

Another proposed scenario for purification of the final state of black hole evaporation consist of postulating that information comes out with the Hawking radiation. This view has been advocated by  't Hooft in \cite{'tHooft:1990fr} and raised to a postulate by Susskind et al. \cite{Susskind:1993if} where one declares that ``there exists a unitary $S$-matrix which describes the evolution from infalling matter to outgoing Hawking-like radiation''. See also Page \cite{Page:1993wv}. Notice that such postulate is not easy to accommodate with the causal structure of Fig. \ref{standard} unless some new physics is invoked so that information of the in-falling modes is somehow `registered' at the BH horizon and sent back out to $\sI^+$.
More precisely, if standard QFT on a curved space-times is assumed to be a valid approximation when the curvature around the black hole horizon is low (for large BHs) then no information on the in-falling modes can leak out the horizon without violating causality. Yet, as argued by Page \cite{Page:1993up}, such leaking of information must be important when the BH is still large and semiclassical. Some peculiar quantum gravity effect must take place at the BH horizon. Further tensions arise when trying to describe the physics from the point of view of freely falling observers who (according to the equivalence principle)  should not feel anything special when crossing the BH horizon (the notion of {\em complementarity} of accounts have been evoked in trying to address this issue \cite{Susskind:1993if}). Hence, the above postulate implies that  quantum gravity effects would be important where they are not expected to be (we will refer to such effects as {\em large quantum gravity effects}). 

The existence of such puzzling {\em large quantum gravity effects} in the present scenario was made manifest by the recent analysis of \cite{Braunstein:2009my, Almheiri:2012rt} where it is explicitly 
shown that (assuming the validity of semiclassical QFT in the vecinity of a large BH) one cannot have information go out of a large BH and across its horizon without a catastrophic violation of the equivalence principle at the BH horizon.  

\begin{figure}[h] \centerline{\hspace{0.5cm} \(
\begin{array}{c}
\psfrag{a}{$a$}\psfrag{b}{$b$}
\psfrag{c}{$c$}
\includegraphics[width=3cm]{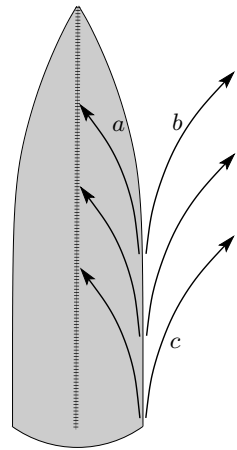}
\end{array}
\)}
\caption{Cartoon representing particle creation in the vicinity of an evaporating BH horizon depicted in Edington-Finkelstein type coordinates. In the complementarity scenario, late particles $b$ are expected to carry non trivial correlations with early particles $c$ as to purify the final state of the Hawking radiation. This view is in contradiction with the equivalence principle and the validity of QFT in the mean field scenario.} \label{AMPS}
\end{figure}

A cartoon description of the phenomenon can be given with the help of Fig. \ref{AMPS} as follows. According to the formalism of QFT on curved space-times the UV structure of the two-point correlation functions is universal for well behaved (Hadamard) states. Physically, this implies that the state of fields looks like  `vacuum' to freely falling observers crossing the horizon with detectors sensitive to wavelengths much shorter that the BH size. In the context of the Hawking effect this implies  that a pair of particles $a$ and $b$ created at the horizon by the interaction of the field with the background geometry must be maximally correlated \cite{Mathur:2009hf}.  The statement that the final state of the Hawking radiation is pure (and thus that information has sneaked out of the horizon during the evaporation process) necessitates the existence of non trivial correlations between the early radiation (particle $c$ in Fig. \ref{AMPS})  and late radiation (particle $b$). But correlations between $c$ and $b$ are forbidden by the fact that $a$ and $b$ are maximally correlated. Relaxing correlations between $a$ and $b$  implies a deviation from the notion of `vacuum' introduced above and the existence of a `firewall' at the horizon, detectable by freely falling observers, created by such excitation when followed back to the past along the horizon. If one is not ready to accept such flagrant violation of the equivalence principle one must admit the inviability of the complementarity scenario.

\subsection{The Ashtekar-Bojowald paradigm}\label{AB}

\begin{figure}[h] \centerline{\hspace{0.5cm} \(
\begin{array}{c}
\psfrag{Ip}{$\sI^{+}$}\psfrag{Im}{$\sI^{-}$}
\psfrag{i0}{$i_0$}\psfrag{ip}{$i^{+}$}\psfrag{im}{$i^{-}$}
\psfrag{s}{$$}\psfrag{S1}{$\Sigma_1$}
\psfrag{S2}{$\Sigma_2$}
\psfrag{u0}{$u_0$}
\psfrag{u1}{$u_1$}
\psfrag{u2}{$u_2$}
\includegraphics[width=5cm]{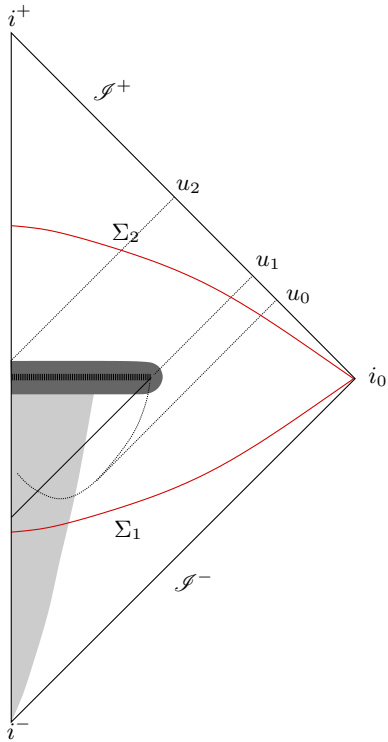}
\end{array}
\)}
\caption{The global space-time causal structure according to the AB-paradigm.
The black hole evaporation takes place according to semiclassical expectations until the
horizon approaches Planck's area. The classical {\em would-be-singularity} is represented by the shaded region where quantum geometry fluctuations are large and no space-time picture is available. The space-time becomes classical to the future of this region: it emerges into a classical (essentially) flat background as required by energy-momentum conservation. Observers at the instant $\Sigma_2$ are in causal contact with the {\em would-be-singularity} which (in classical terms) appears to them as a naked singularity.} \label{ab}
\end{figure}

The means for the resolution of the information puzzle, advocated here, can be formulated in the context of the scenario proposed by Ashtekar and Bojowald (AB) in \cite{Ashtekar:2005cj} whose causal structure is represented in Fig. \ref{ab}. 
The central idea in the latter paper is that the key to the puzzle of information resides in understanding the fate of the classical {\em would-be-singularity} in quantum gravity. This view has enjoyed from a steady consensus in the non perturbative quantum gravity literature \cite{Hossenfelder:2009xq} and it represents the departure guiding principle of the present analysis.  

In the AB proposal the
space-time rapidly becomes semiclassical for an hypothetical observer that initially falls into the black hole, then  enters the high curvature and large geometry-fluctuation dark-shaded region in Fig. \ref{ab},  and finally emerges into a flat space-time future above the {\em would-be-singularity}. The scenario was initially motivated by the observed validity of the unitary evolution across the initial big-bang singularity in symmetry reduced models in the context of {\em loop quantum cosmology} \cite{Bojowald:2001xe} (see \cite{Ashtekar:2011ni} and references therein for a modern account). Similar singularity avoidance results due to the underlying discreteness of LQG  have been reported  recently in the context of spherically symmetric black hole models \cite{Gambini:2014qga}.   The consistency of the AB paradigm is supported by the analysis of \cite{Ashtekar:2008jd} in two dimensional CGHS black holes \cite{PhysRevD.45.R1005} where still some assumptions about the validity of quantum dynamics across the singularity are made.  Numerical investigations of the CGHS model in the mean field approximation \cite{Ashtekar:2010hx, Ashtekar:2010qz} strongly suggest the global causal picture proposed in the AB paradigm as well.

A more detailed description of the dynamical content of Fig. \ref{ab}  will be given in Section \ref{yo} where we will propose the central idea of this work and argue that it leads to a consistent picture of unitary evolution in the BH formation-evaporation process. 

\subsection{Baby Universe}
There is also the possibility of a baby universe, causally disconnected from the assymptotically flat region representing exterior observers, to develop in the future of the {\em would-be-singuarity}. For a discussion of different options and their Penrose diagram representation see \cite{Hossenfelder:2009xq}. Such scenario would be compatible with the perspective developed here; however, for simplicity we will not consider it further in the present discussion.

\section{The argument}

Our argument is based on the assumption that a theory of quantum gravity 
will necessarily imply  a radical change in the way space-time is conceived. 
We are assuming that at the fundamental scale space-time is replaced by a more basic notion made of fundamentally discrete constituents governed by quantum mechanical laws. 

For concreteness we will set the discussion in the context of LQG; however, we believe that the picture presented here
is general enough to resonate with other approaches proposed in the literature.
For instance the {\em group field theory} formulations \cite{Oriti:2014uga}. The {\em asymptotic safety scenario} suggesting a quite different space-time `fractal' picture at the fundamental level with effective dimensional reduction from 4 to 2 \cite{Lauscher:2005qz}. Similar dimensional reduction in {\em dynamical triangulations} have been reported \cite{Ambjorn:2005db}.   Another framework that could be included in the present discussion is the {\em causal sets approach} \cite{Surya:2011yh}.  

A common feature of all these formulations is that space-time arises from a suitable coarse graining where  details of the the relationships among the fundamental pre-geometric building blocks are lost in the limit where the smooth space-time of low energy physics is recovered. It is  reasonable to expect that a prescription of a particular smooth geometry (like flat space-time) will correspond in some of these formulations to infinitely many different fundamental states: an infinite degeneracy of the notion of smooth geometry.
This is the basic assumption that we will make use of in the present argument which is well supported by what is known about the continuum limit in LQG.

\subsection{Smooth space-time from spin-network states in LQG}\label{sst}

In a non perturbative formulation of quantum gravity space-time itself is a dynamical variable to be  quantized and described in the absence of any background reference geometry. In such context, recovering the low energy regime of the theory means to simultaneously recover the field excitations of QFT as well as the smooth space-time geometry where they live and evolve.  In this sense, even in the `simplest' case of 
QFT on Minkowski space-time, the coherent contribution of the infinitely many  underlying fundamental degrees of freedom responsible for the emergence of a definite flat {\em background} space-time must be understood.

In the precise context of LQG the key result along these lines is that space-time geometric operators acquire discrete spectra. States of the gravitational degrees of freedom can be spanned in terms of spin-network states (polymer-like excitations of quantum geometry) each of which admits the interpretation of an eigenstate of geometry which is discrete and atomistic at the fundamental level \cite{Rovelli:1994ge, Ashtekar:1996eg, Ashtekar:1997fb}. The loop representation of the basic algebra of geometric observables has been shown to be the unique one containing a `vacuum' or `no-geometry' state which is diffeomorphism invariant and hence for which all geometric eigenvalues vanish \cite{Lewandowski:2005jk}.  
In this picture flat Minkowski space-time must be viewed as a highly exited state of such `no-geometry' state, where the quantum space-time building blocks are brought together to produce the flat arena where other particles interact. This is a direct implication of the canonical quantization of gravity a la Dirac where the space-time metric becomes a quantum operator
on a Hilbert space. Thus, there is no a priori notion of space-time unless a particular state is chosen in the Hilbert space. 
Loop quantum gravity is a concrete implementation of such non-perturbative canonical quantization of gravity \cite{Rovelli:2004tv, Thiemann:2007zz}.
Even though  important questions remain open, there are robust results exhibiting features which one might expect to be sufficiently generic to remain in a consistent complete picture. 

These states are the boundary data of the quantum theory whose physical content is encoded in  
transition amplitudes to be computed by suitably implementing the quantum constraints stemming from diffeomorphism invariance.  The latter define the physical inner product of solutions \cite{Ashtekar:2004eh} and can be computed via the path integral representation \cite{Perez:2012wv}.
All dynamics is encoded in these amplitudes.
   
There has been an important activity in trying to construct semiclassical states in the framework. At the canonical level efforts have concentrated in the definition of coherent states of quantum geometry \cite{Thiemann:2000bw, Freidel:2010aq, Livine:2007vk, Rovelli:2010km, Bianchi:2010gc} representing a given classical configuration. The relationship between the fundamental spin-network state representation of quantum gravity and the Fock state representation of QFTs has been explored in \cite{Ashtekar:1992tm, Sahlmann:2002qj, Ashtekar:2001xp,  Conrady:2004ww}. 
These results provide useful insights on the nature of the low energy limit of LQG. Unfortunately a final clear understanding is still missing.

However, the view that arises from the above studies is that smooth  geometry should emerge  from the underlying discrete fundamental structures via the introduction of coarse observers insensitive to the details of the UV underlying structures. 
One expects that renormalization group techniques should be essential in such context (see \cite{Dittrich:2014ala} and references therein). The problem remains a hard one as one needs to recover a continuum limit from the underlying purely combinatorial Wheeler's {\em pre-geometric} picture \cite{Misner:1974qy}. The task is complicated further in that the usual tools applied to more standard situations---where continuity arises from fundamentally discrete basic elements (e.g. condensed matter systems or lattice regularisations of QFTs)---cannot be directly imported to the context where no background geometry is available.  Nevertheless, some features seem quite clear from recent studies. In particular any given classical space-time does not correspond to a unique quantum state in the fundamental theory: generically, there will be infinitely many different quantum states satisfying the coarse graining criterion defined by a single classical geometry. For instance, there is no state in the LQG Hilbert space that corresponds to Minkowski space-time, but, rather, to an ensemble of states. In this sense flat space-time is more naturally associated with a density matrix in LQG than to a pure state.  The entropy associated to the latter would be infinite\footnote{Such divergence is an IR one due to the infinite volume of space UV divergences regulated by the presence of a fundamental scale.}.

Unfortunately, due to the difficulties associated with describing the low energy limit of LQG, one cannot give a precise description of the exact nature of the pre-geometric defects that might survive in a state defining a background semiclassical geometry. 
Nevertheless, there are examples of structures that do not seem to play an important role in the construction of the continuum semiclassical space-time, yet they are expected to arise in strong coupling dynamical processes as no selection rule forbids them.  For instance one has {\em close loops} and embedded {\em knots} which are the simplest solutions  to all the quantum constraint-equations \cite{PhysRevLett.61.1155}. As geometric excitations they are degenerate  and carry area  but no volume quantum numbers. It would seem that they could be entangled with the rest of the polymer structure satisfying the  appropriate semiclassical and continuum requirement without affecting it.  More generally the semiclassical weave states in LQG can contain local degenerate contributions such as  {\em trivalent spin-network nodes} or other configurations  with vanishing volume quanta.  On the dynamical side the vertex amplitude of {\em spin foams} \cite{Engle:2007wy} is known not to strongly impose some of the metricity constraints. This allows for the possibility of pre-geometric structures to survive \cite{Speziale:2010cf} in the semiclassical limit. From the canonical  perspective the quantum Hamiltonian constraint seem to invariably produce such pre-geometric defects at the Planck scale. It is reasonable to assume that these should play an important dynamical role when curvature scales become Planckian close to the {\em would-be-singularity} inside black holes. 

The emerging picture in an approach as LQG is certainly very different from the bulk-boundary-duality type of quantum gravity scenarios such as the one proposed by  ADS-CFT correspondence scenarios \cite{Maldacena:1997re}. In LQG the notion of smooth space-time has a capacity for infinite bulk entropy \cite{Livine:2007sy, Ariwahjoedi:2014wpa}. Such non holographic behaviour at the fundamental level might seem at odds with the phenomenology behind the belief in the so-called {\em holographic principle} as a basic pillar of quantum gravity  \cite{Bousso:2002ju}.  However, further scrutiny shows that predictions of the LQG framework are completely consistent with the holographic phenomenology. 

This is basically because, in the physical situations where holographic behaviour arises, bulk entropy only contributes as an irrelevant overall constant.  A clear example of this is the validity of covariant entropy bounds satisfied by relative (entanglement) entropy \cite{Bousso:2014sda}. Relative entropy of a state $\rho$ is defined with respect to a reference quantum state (a `vacuum' state) $\rho_0$ as
\be\label{ree}
S_{\rho_0}(\rho) = - \Tr[\rho \log \rho] + \Tr[\rho_0 \log \rho_0].
\ee
While giving non trivial information about excitations of the `vacuum' in the mean field approximation (where a space-time background makes sense) such covariant entropy bounds do not constraint the number of fundamental degrees of freedom of quantum geometry. More precisely, the bulk Planckian entropy, being a constant, just cancels out in the subtraction that regularises $S_{\rho_0}(\rho)$. The point here is that on the basis of the insights of LQG on the nature of the  fundamental quantum geometry excitations, the holographic principle is degraded from its status of fundamental principle to a low energy property of quantum fields on curved space-times. This is the perspective that follows from the LQG statistical mechanical account of black hole entropy (see \cite{Perez:2014ura} for a description in terms of entanglement and a discussion along the lines of the present paper and \cite{DiazPolo:2011np, bayo} for recent reviews). In this way the framework of LQG, without being fundamentally `holographic', can accommodate the holographic phenomenon when it is valid. The holographic behaviour is a characteristic of suitable systems (black holes, isolated horizons, null surfaces, etc.) but not a fundamental property of the theory.

Specialising to ensemble of states that all `look like' Minkowski space-time for suitably defined coarse grained observers we notice that their members must differ by hidden degrees of freedom from the viewpoint of those low energy observers. In particular they would all agree on stating that all the states (even though different at the fundamental scale) have zero ADM (or Bondi) energy. We have seen above that in the particular case of LQG a whole variety of pre-geometric structures, that are well characterized in the strong coupling regime, are potential local defects that, as they do not affect the flatness of the geometry of the semiclassical state would carry no energy. These Planckian defects in the fabric of space-time are hidden to the low energy observers but represent genuine degrees of freedom to which other degrees of freedom can correlate to.  The assumption that such structures are there is the core of the proposed paradigm for understanding the fate of information in black hole evaporation.

\subsection{Revisiting the AB paradigm}\label{yo}

The discussion of the previous sections suggest a natural extention of the Ashtekar-Bojowald paradigm evoked in Section \ref{AB}. As we will see now, this extension provides a natural resolution of the information puzzle without invoking {\em large quantum effects} in space-time regions expected to behave classically (e.g. firewalls \cite{Almheiri:2012rt}, fireworks \cite{Rovelli:2014cta, Haggard:2014rza}).  We will argue that the emerging global picture for black hole formation and evaporation is, at the same time,
completely unitary while, due to decoherence with pre-geometric Planckian structures, compatible with the original implications of Hawking's initial statement. We assume the global causal structure of the space-time is the one proposed by Ashtekar  and Bojowald \cite{Ashtekar:2005cj}, see Fig. \ref{ab}. We will also see that some information on the geometry of the {\em would-be-singularity} might be extracted from our argument.

The key ingredient is contained in the previous section: in a theory like LQG, where space-time itself is emergent in the semiclassical limit from a fundamental granular structure involving coarse graining, non trivial information can be encoded with no energy cost. Such degrees of freedom are hidden to standard QFT observers as they are only affected by low energy field-like excitations while they remain insensitive to the UV structure of space-time (an example of such an QFT observable---of central importance in the present discussion---is the notion of relative entanglement entropy (\ref{ree})).  
Loss of quantum coherence with such degrees of freedom explains in this scenario the apparent paradoxes associated to the fate of information in black hole evaporation. 

Assume that we start from a pure quantum state of matter fields and geometry on $\sI^-$ in Fig. \ref{ab}. 
The state is assumed to be a semiclassical state. For that reason the `fast' dynamical process leading to the formation of the BH  is well approximated by the classical field equations. 
The system undergoes gravitational collapse, matter and gravity waves fall into the singularity while matter and gravitational radiation is sent out to $\sI^+$. The `fast' era ends with a stationary BH with Bondi mass $M$ at $u_0$ (we assume the both angular momentum and charge vanish for simplicity). Such classical picture is incomplete in to aspects: the classical singularity inside the horizon indicates the need of a full quantum gravity treatment there, and non-trivial local correlations in the initial state across the horizon lead to the evaporation of the initial Bondi mass as Hawking radiation (a slow quantum gravity effect driving the asymptotic global structure of the space-time in Fig. \ref{ab}).  
%
%

Let us first concentrate on the strong quantum effects close to the singularity. Here we follow \cite{Ashtekar:2005cj}. Close to the singularity the curvature scale approaches Planck's scale and dynamics has to be described by a quantum gravity theory.  The shaded region in Fig \ref{ab} surounding the {\em would-be-singularity} is a region where the usual notion of space-time is inapplicable due to large quantum fluctuations of geometric observables: the fundamental granular structure of the space-time foam precludes the use of any notion of background geometry there. Nevertheless, a unitary evolution of fields and geometric degrees of freedom across this region is expected to be well defined. Moreover, one assumes that as a consequence of this evolution the space-time becomes classical again at some later `time' $\Sigma_2$ to the future of the {\em would-be-singularity}. As mentioned above this assumption is realised in quantum-cosmology as well as in spherical collapse models and further explored in 2d models (see discussion and references in Section \ref{AB}).

If there is no singularity can we talk about a BH horizon in the scenario of Fig. \ref{ab}? At first sight the answer is no. Notice that the conformal diagram  suggests that every space-time point is, according to the discussion above, causally connected with $\sI^+$. Therefore,  there is no BH region in the usual sense. Nevertheless, one can still define the BH region as the complement of all space-time points that can be joined to $\sI^+$ by causal curves that do not enter into the {\em would-be-singularity} region. Basically, here one is using the usual definition with a qualification in the definition of the past of $\sI^+$. We could call this the {\em classical past} of $\sI^+.$ Notice that this is equivalent to using the standard definition in a space-time where the shaded region around the {\em would-be-singularity} is removed from the manifold. Hence, under the usual assumptions of asymptotic predictability and energy conditions the usual (classical) BH theorems would apply (and be physically relevant in the past where quantum effects are still unimportant).  The apparent horizon---a {\em generalized dynamical horizon} represented here by the dotted line---has interesting physical properties that we will not discuss here \cite{AKLR}.

Now we are ready to describe the slow quantum effect of Hawking evaporation from the viewpoint of external observers. Physically, outside of the BH region and after a sufficiently long time $u_0$ (measured in terms of retarded time $u$ at $\sI^+$ see Fig. \ref{ab})  the BH becomes stationary and its state is labelled by an `initial'  Bondi mass $M$. A very slow process of Hawking radiation (described at first well by semiclassical equations) becomes the main dynamical process outside of the BH horizon. The Bondi mass $M_B(u)$ decreases slowly with a rate proportional to $\ell_p^2/M^2_B(u)$ during the semiclassical era. When the Bondi mass of the BH approaches the Planck mass a full quantum gravity treatment is necessary. The {\em would-be-singularity} becomes naked at $u_1$. From $u_1$ to $u_2$ the {\em would-be-singularity} is visible from $\sI^+$. By the time $u_2$ the Bondi mass has become basically zero, the space-time is classical again and hence isomorphic to a portion of Minkowski space-time.  

The causal structure in Fig. \ref{ab}, which in a loose sense is globally hyperbolic (yet recall that no space-time interpretation is possible in the grey region around the {\em would-be-singularity}), and the assumption of unitarity of the underlying quantum theory implies that 
the information contained in an initial state defined at instant $\Sigma_1$ should be recovered in the state at instant $\Sigma_2$.  If the system  is to be unitary, the causal structure tells one that the purification of the radiation from the BH evaporation must take place with degrees of freedom excited after time $u_1$. This by itself eliminates the {\em fire wall} problem but it raises the question on how the information can leave after $u_1$. 
The BH {\em inside}-{\em outside} correlations at $\Sigma_1$ should become correlations between the whole of the early radiation due to the collapse process plus the Hawking radiation---both emitted before $u_1$---and {\em something} encoded at $\sI^+$ after $u_1$. Most importantly, as pointed out in the discussion about remnants, the purifying excitations after $u_1$ must carry 
basically zero energy.  

\begin{figure}[h] \centerline{\hspace{0.5cm} \(
\begin{array}{c}
\psfrag{s1}{$\Sigma$}\psfrag{b}{$b$}\psfrag{a}{$a$}\psfrag{ab}{$\bar a$}
\psfrag{s2}{$\Sigma^{\prime}$}
\includegraphics[width=6cm]{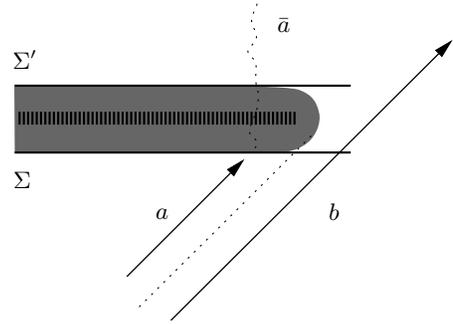}
\end{array}
\)}
\caption{Two instants of `time' before and after the {\em would-be-singularity}.
The spacial surface $\Sigma$ is a latest surface where the space-time notion is still applicable. The surface $\Sigma^{\prime}$ is the earliest space-like surface in the flat emerging flat space-time across the {\em would-be-singularity}. The particles $a$ and $b$ are created close to the BH horizon. 
 Particle $b$ escapes to infinity as Hawking radiation. Particle $a$ falls into the singularity, deposits its negative energy load, striped off its energy it emerges unitarily transformed into a defect $\bar a$ in the quantum weave state describing flat space-time to the future of the {\em would-be-singularity}.} \label{slicing}
\end{figure}

What is the nature of the purifying degrees of freedom? In order to answer this question let us first consider the heuristic picture of pair creation at the BH horizon\footnote{Even when such mental representation of the Hawking process is quite simplistic, it will allow for a clear statement of the idea proposed here. Any concrete calculation would require a careful definition of the mathematical setup including a full background independent treatment of quantum geometry. An attempt to go along such line here will only obscure the main point.} as depicted in Fig. \ref{slicing} which is a zoom of the {\em would-be-singularity} region of Fig. \ref{ab}. A pair $a$-$b$ is `created' and particle $b$ escapes to infinity to be part of Hawking radiation. The positive energy flux of particles like $b$ at $\sI^+$ and the semiclassical field equations imply the decrease of the Bondi mass mentioned above.  While particle $b$ carries positive energy to $\sI^+$ particle $a$ carries `negative energy' to the singularity \footnote{Such energy is not energy for any local observer, it is just the conserved quantity associated to the stationarity Killing field that is space-like inside when the BH evaporates slowly. Notice however that as the particle falls into the singularity, even in the region wether space-time is semiclassical, it will not necessarily be well approximated by the spherically symmetric solution. This is due to the fact that all the  complexity of the initial conditions leading to the collapse are visible inside. If one would take would take the spherical background model all the way to the singularity then a particle $a$ with zero angular momentum has a local energy vanishes when it approaches the singularity.}.

Now particle $a$ falls into the region where the semiclassical description is no longer applicable.
The conservation of energy arguments at null infinity given above require $a$ to be striped off its energy and to emerge on the future of the singularity transformed into a new degree of freedom $\bar a$ (which we shall argue is a pre-geometry defect of the kind described in Section \ref{sst}). The conservation of energy argument at future null infinity is global. How can one understand this process locally. The net energy-momentum flux across the boundary of the shaded region in Fig. \ref{ab} (around the {\em would-be-singularity}) is clearly not zero. This is of course not a problematic issue: On the one hand,  local conservation of energy momentum should not hold as the geometry there (as long as the semiclassical approximation makes sense) is extremely dynamical. On the other hand, the semiclassical approximation is not good close to the {\em would-be-singularity} so the very notion of energy-momentum (tied to a notion of background space-time) becomes ill defined there. 

\begin{figure}[h] \centerline{\hspace{0.5cm} \(
\begin{array}{c}
\psfrag{Ip}{$\sI^{+}$}\psfrag{Im}{$v=0$}
\psfrag{i0}{$i_0$}\psfrag{ip}{$i^{+}$}\psfrag{im}{$i^{-}$}
\psfrag{a}{$ds^2=-du dv$}
\psfrag{b}{$ds^2=-({\van 1-\frac{2M}{r}}) du_s dv$}
\psfrag{S1}{$\Sigma_1$}
\psfrag{S2}{$\Sigma_2$}
\psfrag{s}{$$}
\includegraphics[width=6cm]{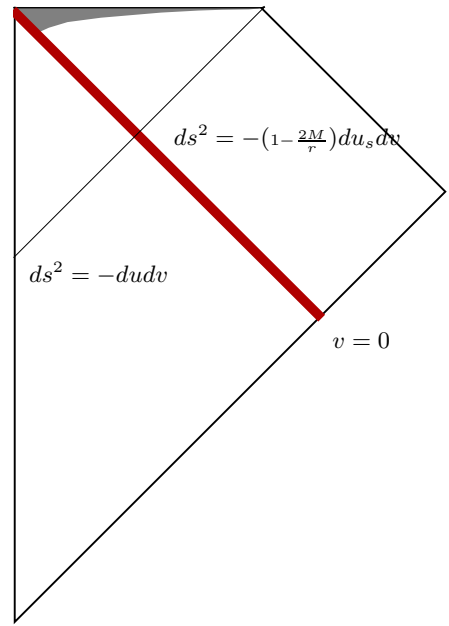}
\end{array}
\)}
\caption{2d spherical black hole made from the gravitational collapse of a spherical pulse of energy $M$.
The metric is flat inside the shell and Schwarzschild outside. Continuity of the metric across the shell implies
the following relationship between retarded time $u=t-r$ and $u_s=t-r_*$ (for $r_*$ the standard tortoise coordinate):
$u_s=u-4M\log(1+\frac{u}{4M})$ and $v_s=v$. Coordinates are chosen so that the shell collapse takes place at $v=0$. The expectation value of the energy momentum tensor in the Unruh vacuum is known in close form \cite{Fabbri:2005mw} everywhere in the spacetime. The shaded area denotes qualitatively the region where observers falling along $\partial_r$ detect energy densities smaller than some negative fixed value.} \label{ccc}
\end{figure}

We would like to add a discussion involving the expectation value of the 
energy momentum tensor inside the black hole region in the suitable quantum states representing the physical situation at hand.  This analysis conforms very well with the scenario presented here. To simplify the treatment, and make an analytic calculation possible, we concentrate on $s$-modes and further neglect backscattering effects. Under such assumptions one can describe the qualitative features of the semiclassical collapse in an 2d effective space-time. For concreteness we model the spherically symmetric collapse by a shock 
wave of (ADM) mass $M$ (see Figure \ref{ccc}). The Unruh vacuum \cite{Unruh:1976db} describes well the physics that one has in mind in the present context: the state in the past is the Minkowski vacuum for out going modes and there are no incoming particles from $\sI^-$. The  expectation value of all components of $T_{\mu\nu}$ in that state can be computed explicitly \cite{Davies:1976ei} (see Chapter 5 of \cite{Fabbri:2005mw} for a textbook treatment). For concreteness, if we consider freely falling (zero angular momentum) observers with four velocities $u^a$ then it follows that close to the singularity one gets
\be\label{timunu}
\langle T_{ab} u^au^b \rangle\approx -\frac{\ell_p^2 M}{48 r^5}\left[1+\left(\frac{r}{u}\right)^4\right],
\ee 
where $u=r-t$ in terms of the Minkowski spherical coordinates defined inside the collapsing shell so that $t=0$ and $r=0$
corresponds to the event where the shell shrinks to zero (the previous expression is valid to the future of the shell where the energy-momentum tensor is non trivial). One sees how the weak energy condition is strongly violated by quantum effects 
as these observers approach the singularity\footnote{The naturalness of the appearance of negative energies in the strong curvature regime follows directly from the 
quantum inequalities put forward in \cite{Flanagan:2002bd}. In fact (\ref{timunu}) saturates the bound when $r=u$, i.e., the point where the shell meets the singularity in Figure \ref{ccc}.}. In particular, the second term makes negative energies become more important close to the event where the shell meets 
the singularity. 
The previous expression for the energy momentum tensor (close to and) on the past of the {\em would-be-singularity} is a first hint of a possible dynamical explanation of the 
flattening the space-time to the future of the  {\em would-be-singularity} as the annihilation of the positive energy $M$ by a $\langle T_{ab}\rangle$ that strongly violates energy conditions there.
Of course the above argument can only be taken as an indication in the right direction. 
A serious study of this issue requires a non perturbative analysis\footnote{Notice that such contributions to the energy-momentum tensor comes from the infinitely many degrees of freedom of quantum fields on the geometry. By definition such contributions are not taken into account in the symmetry reduced models predicting essentially a `time-symmetric' bounce around the {\em would-be-singularity} \cite{Ashtekar:2011ni}. The same remark can be made to the recently introduced spherically symmetric quantum collapse models \cite{Gambini:2014qga} where matter fields are not included (the $s$-modes leading to (\ref{timunu})).  We believe that, until local excitations are suitably taken into account, the `time-symmetric' bounce predicted by symmetry reduced models cannot be taken too literally.}.  

The transmutation of a QFT degree of freedom $a$ into a pre-geometric one $\bar a$ seem possible even from very general classical and quantum arguments. For instance, according to the BKL conjecture \cite{BKL}
as we approach the space-like singularity individual points decouple and geometry evolves in a chaotic manner
which is well understood in certain cosmological scenarios \cite{lrr-2008-1}. 
Such ultra-local dynamics can formally be recovered in the strong coupling regime by simple dimensional analysis. 
These analysis have been used to argue for the existence of a dimensional reduction from 4d to 2d as one approaches the UV regime in quantum gravity. Such picture is in agreement with similar results obtained in various approaches to quantum gravity and fully compatible with our discussion of Section \ref{sst} (see \cite{Carlip:2009km} and references therein). In such context the nature of the field degrees of freedom $a$ in the Planckian environment  of the {\em would-be-singularity} should be expected to be very different from its original Fock-like excitation at low energies.
 
Unitarity implies that the correlations between the degrees of freedom of $\bar a$ and $b$ remain intact.  Therefore,  
 $\bar a$ must carry no energy but yet correspond to true degrees of freedom. These cannot be propagating degrees of  freedom on the flat background after evaporation because as such they would have to carry energy. Then one would be back to a scenario similar to that of a long lived remnant and with its standard difficulties. However, as argued in Section \ref{sst} there is an infinite pool of Planckian degrees of freedom in background independent formulations of quantum gravity. Moreover, there seem not to be any selection rule protecting these defects from developing\footnote{In certain simplified toy dynamical models for black hole evaporation,  creation of such defects has been argued to take place from the generic form of the action of the Hamiltonian constraint \cite{Pranzetti:2012dd}.}. With the view that anything that is allow to happen will happen to preserve the unitarity of the underlying fundamental description---and in the context of the space-time Fig \ref{ab}---$\bar a$  must be one of these pre-geometry   defects in the underlying space-time weave hidden to low energy observers but yet carrying non trivial information.

\begin{figure}[h] \centerline{\hspace{0.5cm} \(
\begin{array}{c}
\psfrag{a}{$a$}\psfrag{b}{$b$}
\psfrag{ab}{$\bar a$}
\psfrag{c}{$c$}
\includegraphics[width=3cm]{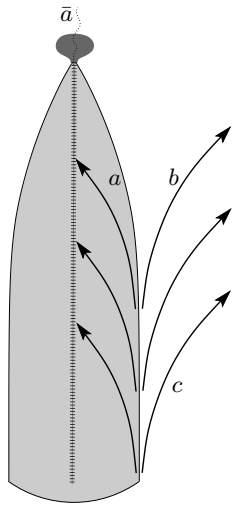}
\end{array}
\)}
\caption{In a theory of quantum gravity, where fundamental geometric excitatoins are discrete and smooth space-time arises from coarse graining, there are infinitely many degrees of freedom that might store information with no energy cost.
In such scenario the mixed state representing the Hawking radiation from the evaporated BH can be purified by correlations with the hidden degrees of freedom becoming in causal contact with future observers once the {\em would-be-singularity} becomes visible. Particles $b$ and $c$ are completely uncorrelated as expected from standard QFT considerations. Particle $a$ emerges as a defect $\bar a$ from the naked singularity while remaining maximally correlated with $b$.} \label{defects}
\end{figure}

In this scenario BH evaporation ends with an extended space-time event where the {\em would-be-singularity} becomes 
visible to the outside world. To the future of that event (for retarded asymptotic time  $u>u_1$) the pre-geometric purifying degree of freedom $\bar a$ establish causal contact with the outside. They emerge from the naked {\em would-be-singularity} region, where fluctuations of geometry are so high that no space-time picture is allowed, as illustrated in Fig. \ref{defects}.

\begin{figure}[h] \centerline{\hspace{0.5cm} \(
\begin{array}{c}
\psfrag{s1}{$\Sigma$}\psfrag{b}{$b$}
\psfrag{s2}{$\Sigma^{\prime}$}
\includegraphics[width=6cm]{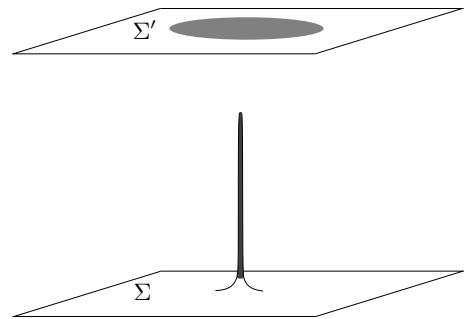}
\end{array}
\)}
\caption{A qualitative representation of the Riemannian geometry of $\Sigma$ and $\Sigma^{\prime}$ of Fig \ref{slicing}. The shaded regions are those `touching' the quantum region.} \label{slicing2}
\end{figure}

In order to get some intuition on what the geometry of the naked {\em would-be-singularity} region might be, we present the following heuristic argument. This argument cannot be taken too literally as the only clear geometric picture must come from a dynamical description using the full quantum gravity machinery.
In Fig. \ref{slicing2} one is representing the geometry of the two space-like hypersurfaces $\Sigma$ and $\Sigma^{\prime}$ of Fig. \ref{slicing}. The volume of the shaded region in $\Sigma$ depends on the details of the evaporation process and on quantum gravity effects in the non-semiclassical regime. From dimensional arguments and in the large mass $M$ regime it goes like
\be V(\Sigma)\propto M^3 ({M}/{\ell_p})^{\alpha}.\ee  
where the missing proportionality constant and $\alpha$ depend on the interior dynamics.
For instance one gets $\alpha=5/2$ if one (toy-)models the evaporation process with an advanced Vaidya metric.  
We can estimate the scaling of the volume of the shaded region in the flat hypersurface $\Sigma^{\prime}$ by requiring that there are at least as many volume `bits' as necessary to purify the radiation emitted during the Hawking evaporation process. This number is of the order of the initial BH entropy $A(M)/(4\ell_p^2)$, where $A(M)$ is the area of the BH at retarded time $u_0$ (see Fig. \ref{ab}). From this one gets
\be V(\Sigma')\propto \ell_p {M}^{2}.\ee  
From which a characteristic size $L= 10^{-11} (M/M_{\odot})^{2/3}m$ of the naked {\em would-be-singularity} follows. For a BH  with $M=10^{15}g$ (e.g.  primordial BHs completing evaporation at present) this gives a size of about $10^{-2} \ell_{LHC}$ where $\ell_{LHC}$ is the shortest scale to which the {\em Large Hadron Collider} is sensitive today.
The life time $\tau=u_2-u_1$ of the naked region as measured from $\sI^+$ would be of the same order in geometric units. So even when the {\em would-be-singularity} region is macroscopic in comparison to the fundamental scale the scale of the naked event can be quite small in cosmological scales. 

\section{Discussion}

The basic idea of this article is that, in a theory of quantum gravity where the fundamental degrees of freedom are discrete and combinatorial at the Planck scale, the effective low energy description in terms of quantum fields living on a smooth background geometry cannot be unitary when singularities in the geometry develop. The lack of unitarity is a consequence of such effective description: close to the {\em would-be-singularity} the smooth background geometry notion breaks down and correlations with Wheeler's pre-geometric structures become important. Unitarity of the fundamental theory implies that  correlations subsist in the future when the state admits again background space-time geometry interpretation.  However pre-geometric structures are now only detectable to low energy QFT observers though the loss of quantum coherence of their effective field variables.  

The notion of smooth geometry is a highly degenerate notion in such a framework: one semiclassical state associated to a space-time $(\sM, g_{ab})$ corresponds to an infinite ensemble of fundamental quantum states differing by pre-geometric local exitations.
In the picture presented here, and at the fundamental quantum gravitational level where evolution is unitary, the initial flat Minkwoski space-time, represented by $\sI^-$,  differs from the final 
flat space-time $\sI^+$ in terms of Planckian defects inflicted in the quantum geometry by the unitary evolution across the would be singularity. The two are, however, indistinguishable for coarse grained low energy standard QFT observers.
In such low energy context is natural to consider geometry and matter as represented by mixed states of the fundamental theory and evolution as given by Hawking's super-scattering operator. However, the full dynamical description in the background independent setting will remain unitary. {This means that information could be retrieved but only via observations that test the fundamental length. Such observations are beyond the regime of applicability of the effective field theory description; they might be precluded in practice but they are allowed in principle.}

In the usual situations where no singularities develop consistency requires that decoherence with the Planckian substructures evoked in our argument to be negligible in order to recover the usual unitarity of standard quantum field theory. This is a non trivial constraint that must be fulfilled by the low energy limit of theories like LQG which in our view is related to the non trivial requirement is Lorentz invariance.  Lorentz invariance implies that  the space-time granularity of quantum geometry cannot be interpreted literally as associated to a special rest frame \cite{Collins:2004bp, Collins:2006bw, Polchinski:2011za} or in purely classical terms \cite{Rovelli:2002vp}.    Space-time granularity compatible with Lorentz invariance becomes important only in regimes of high curvature \cite{Corichi:2005fw, Aguilar:2012ju}. It is in these situations where decoherence between QFT degrees of freedom and defects in the space-time weave becomes important. 

The picture presented is very much compatible with the second law.
Close to the {\em would-be-singularity} space-time granularity becomes relevant and so new degrees of freedom become available. Information is not lost at the fundamental level where unitarity holds; however, the unitary dynamics
entangles the low energy degrees of freedom (field excitations on a background geometry) with those quantum geometric degrees of freedom that do not admit such low energy characterisation. Correlations with these hidden variables   become important and information in QFT type of variables is degraded by decoherence. 
The quantum `defects' created in the `fire' of the high curvature region close to the singularity  remain hidden to external observers for a long time of order $M^3/\ell_p^2$ but regain causal contact with them when the BH completely evaporates its energy content in Hawking radiation. These `defects' are like 
the ashes of the naked singularity that purify the final state of the whole system.

{ Finally, our scenario does not require any correlations between the early and late Hawking radiation for $u\in [u_0,u_1]$ in Figure \ref{ab}. This means that the quantum state of fields during the semiclassical era are allowed to be maximally correlated across the horizon as there is no need to evacuate information to preserve unitarity during the Hawking evaporation process. More precisely, to the past of a sufficiently early Cauchy surface before the {\em would-be-singularity}---where a formulation in terms of standard QFT on curved space-time is a good approximation---the quantum state of fields on the space-time representing gravitational collapse must be a Hadarmard state with its characteristic short length correlations, in particular across the horizon. For such states, the expectation value of the energy-momentum tensor will be regular at the horizon which implies the absence of any firewalls. }
  
The present scenario was presented in an earlier form in \cite{ffp14}.  We discovered after that the idea of invoking degrees of freedom carrying no energy as key for understanding the BH evaporation has  been advocated before by  
Unruh and Wald  \cite{Unruh:1995gn} and more recently by Unruh \cite{Unruh:2012vd}. As mentioned in the introduction the first reference deals with  the apparent energy-momentum conservation violations associated with Hawking's super-scattering formalism showing that they disappear if one involves decoherence with energy cheap degrees of freedom.
Along these lines Unruh \cite{Unruh:2012vd}  studies a simple example of a scattering process where non trivial entanglement with a spin system takes place. He uses this example to argue for a perspective leading to a possible resolution of information loss  issue in BH evaporation. As pointed out by Unruh condense matter systems  exhibiting this type of decoherence exist (Unruh provides some references for this, in particular \cite{PhysRevLett.97.207206}). The perspective explored in the present paper goes along similar lines. Its novelty resides in pointing at a concrete candidate for these hidden degrees of freedom while providing a space-time causal representation of the evaporation process.  

\subsection{Acknowledgements}

I am grateful to I. Agullo, F. Barbero, J. Pullin and D. Sudarsky for useful interactions at the beginning of this project. 
I also thank discussions with E. Bianchi,  H. Haggard,  A. Riello, D. Pranzetti,  C. Rovelli, and Simone Speziale. This work has been carried with partial support from the OCEVU Labex (ANR-11-LABX-0060) and the A*MIDEX project (ANR-11-IDEX-0001-02) funded by the ``Investissements d'Avenir" French government program managed by the ANR.

\end{document}